\documentclass[twocolumn,prd,showpacs]{revtex4}
\usepackage{graphicx}
\begin{document}
\def\Journal#1#2#3#4{{#1} {\bf #2}, #3 (#4)}

\def\NCA{Nuovo Cimento}
\def\NIM{Nucl. Instrum. Methods}
\def\NIMA{{Nucl. Instrum. Methods} A}
\def\NPB{{Nucl. Phys.} B}
\def\NPBs{{Nucl. Phys.} B (Proc. Suppl.)}
\def\PLB{{Phys. Lett.}  B}
\def\PRL{Phys. Rev. Lett.}
\def\PRD{{Phys. Rev.} D}
\def\ZPC{{Z. Phys.} C}
\def\APJ{Astrophys. J.}
\def\SNP{Sov. Jour. Nucl. Phys.}

\newcommand{\plumin}[2]{^{+#1}_{-#2}}
\newcommand{\plum}[2]{\matrix{+#1\\[-3pt]-#2}}
\newcommand{\plumi}[2]{\matrix{+#1\\[-1.2mm]-#2}}

\def\dmsq{\Delta m^2}
\def\tasq{\tan^2\theta}

\title{Precise Measurement of the Solar
Neutrino Day/Night and Seasonal Variation in Super-Kamiokande-I}

\newcounter{foots}
\newcounter{notes}
\newcommand{\authoraticrr}{$^{1}$}
\newcommand{\authoratncen}{$^{2}$}
\newcommand{\authoratbu}{$^{3}$}
\newcommand{\authoratbnl}{$^{4}$}
\newcommand{\authoratuci}{$^{5}$}
\newcommand{\authoratcsu}{$^{6}$}
\newcommand{\authoratcnu}{$^{7}$}
\newcommand{\authoratgmu}{$^{8}$}
\newcommand{\authoratgifu}{$^{9}$}
\newcommand{\authoratuh}{$^{10}$}
\newcommand{\authoratkek}{$^{11}$}
\newcommand{\authoratkobe}{$^{12}$}
\newcommand{\authoratkyoto}{$^{13}$}
\newcommand{\authoratlanl}{$^{14}$}
\newcommand{\authoratlsu}{$^{15}$}
\newcommand{\authoratumd}{$^{16}$}
\newcommand{\authoratmit}{$^{17}$}
\newcommand{\authoratduluth}{$^{18}$}
\newcommand{\authoratsuny}{$^{19}$}
\newcommand{\authoratnagoya}{$^{20}$}
\newcommand{\authoratniigata}{$^{21}$}
\newcommand{\authoratosaka}{$^{22}$}
\newcommand{\authoratseoul}{$^{23}$}
\newcommand{\authoratshizuokaseika}{$^{24}$}
\newcommand{\authoratshizuoka}{$^{25}$}
\newcommand{\authoratskku}{$^{26}$}
\newcommand{\authorattohoku}{$^{27}$}
\newcommand{\authorattokyo}{$^{28}$}
\newcommand{\authorattokai}{$^{29}$}
\newcommand{\authorattit}{$^{30}$}
\newcommand{\authoratwarsaw}{$^{31}$}
\newcommand{\authoratuw}{$^{32}$}

\newcommand{\addressoficrr}[1]{$^{1}$ #1 }
\newcommand{\addressofncen}[1]{$^{2}$ #1 }
\newcommand{\addressofbu}[1]{$^{3}$ #1 }
\newcommand{\addressofbnl}[1]{$^{4}$ #1 }
\newcommand{\addressofuci}[1]{$^{5}$ #1 }
\newcommand{\addressofcsu}[1]{$^{6}$ #1 }
\newcommand{\addressofcnu}[1]{$^{7}$ #1 }
\newcommand{\addressofgmu}[1]{$^{8}$ #1 }
\newcommand{\addressofgifu}[1]{$^{9}$ #1 }
\newcommand{\addressofuh}[1]{$^{10}$ #1 }
\newcommand{\addressofkek}[1]{$^{11}$ #1 }
\newcommand{\addressofkobe}[1]{$^{12}$ #1 }
\newcommand{\addressofkyoto}[1]{$^{13}$ #1 }
\newcommand{\addressoflanl}[1]{$^{14}$ #1 }
\newcommand{\addressoflsu}[1]{$^{15}$ #1 }
\newcommand{\addressofumd}[1]{$^{16}$ #1 }
\newcommand{\addressofmit}[1]{$^{17}$ #1 }
\newcommand{\addressofduluth}[1]{$^{18}$ #1 }
\newcommand{\addressofsuny}[1]{$^{19}$ #1 }
\newcommand{\addressofnagoya}[1]{$^{20}$ #1 }
\newcommand{\addressofniigata}[1]{$^{21}$ #1 }
\newcommand{\addressofosaka}[1]{$^{22}$ #1 }
\newcommand{\addressofseoul}[1]{$^{23}$ #1 }
\newcommand{\addressofshizuokaseika}[1]{$^{24}$ #1 }
\newcommand{\addressofshizuoka}[1]{$^{25}$ #1 }
\newcommand{\addressofskku}[1]{$^{26}$ #1 }
\newcommand{\addressoftohoku}[1]{$^{27}$ #1 }
\newcommand{\addressoftokyo}[1]{$^{28}$ #1 }
\newcommand{\addressoftokai}[1]{$^{29}$ #1 }
\newcommand{\addressoftit}[1]{$^{30}$ #1 }
\newcommand{\addressofwarsaw}[1]{$^{31}$ #1 }
\newcommand{\addressofuw}[1]{$^{32}$ #1 }

\author{
{\large The Super-Kamiokande Collaboration} \\ 
\bigskip
M.B.~Smy\authoratuci,
Y.~Ashie\authoraticrr,
S.~Fukuda\authoraticrr,
Y.~Fukuda\authoraticrr,
K.~Ishihara\authoraticrr,
Y.~Itow\authoraticrr,
Y.~Koshio\authoraticrr,
A.~Minamino\authoraticrr,
M.~Miura\authoraticrr,
S.~Moriyama\authoraticrr
% ...
\\
\smallskip
\footnotesize
\it
\addressoficrr{Kamioka Observatory, Institute for Cosmic Ray Research, University of Tokyo, Kamioka, Gifu, 506-1205, Japan}\\
\addressofncen{Research Center for Cosmic Neutrinos, Institute for Cosmic Ray Research, University of Tokyo, Kashiwa, Chiba 277-8582, Japan}\\
\addressofbu{Department of Physics, Boston University, Boston, MA 02215, USA}\\
\addressofbnl{Physics Department, Brookhaven National Laboratory, Upton, NY 11973, USA}\\
\addressofuci{Department of Physics and Astronomy, University of California, Irvine, Irvine, CA 92697-4575, USA }\\
\addressofcsu{Department of Physics, California State University, Dominguez Hills, Carson, CA 90747, USA}\\
\addressofcnu{Department of Physics, Chonnam National University, Kwangju 500-757, Korea}\\
\addressofgmu{Department of Physics, George Mason University, Fairfax, VA 22030, USA }\\
\addressofgifu{Department of Physics, Gifu University, Gifu, Gifu 501-1193, Japan}\\
\addressofuh{Department of Physics and Astronomy, University of Hawaii, Honolulu, HI 96822, USA}\\
\addressofkek{Institute of Particle and Nuclear Studies, High Energy Accelerator Research Organization (KEK), Tsukuba, Ibaraki 305-0801, Japan }\\
\addressofkobe{Department of Physics, Kobe University, Kobe, Hyogo 657-8501, Japan}\\
\addressofkyoto{Department of Physics, Kyoto University, Kyoto 606-8502, Japan}\\
\addressoflanl{Physics Division, P-23, Los Alamos National Laboratory, Los Alamos, NM 87544, USA }\\
\addressoflsu{Department of Physics and Astronomy, Louisiana State University, Baton Rouge, LA 70803, USA }\\
\addressofumd{Department of Physics, University of Maryland, College Park, MD 20742, USA }\\
\addressofmit{Department of Physics, Massachusetts Institute of Technology, Cambridge, MA 02139, USA}\\
\addressofduluth{Department of Physics, University of Minnesota, Duluth, MN 55812-2496, USA}\\
\addressofsuny{Department of Physics and Astronomy, State University of New York, Stony Brook, NY 11794-3800, USA}\\
\addressofnagoya{Department of Physics, Nagoya University, Nagoya, Aichi 464-8602, Japan}\\
\addressofniigata{Department of Physics, Niigata University, Niigata, Niigata 950-2181, Japan }\\
\addressofosaka{Department of Physics, Osaka University, Toyonaka, Osaka 560-0043, Japan}\\
\addressofseoul{Department of Physics, Seoul National University, Seoul 151-742, Korea}\\
\addressofshizuokaseika{International and Cultural Studies, Shizuoka Seika College, Yaizu, Shizuoka, 425-8611, Japan}\\
\addressofshizuoka{Department of Systems Engineering, Shizuoka University, Hamamatsu, Shizuoka 432-8561, Japan}\\
\addressofskku{Department of Physics, Sungkyunkwan University, Suwon 440-746, Korea}\\
\addressoftohoku{Research Center for Neutrino Science, Tohoku University, Sendai, Miyagi 980-8578, Japan}\\
\addressoftokyo{The University of Tokyo, Tokyo 113-0033, Japan }\\
\addressoftokai{Department of Physics, Tokai University, Hiratsuka, Kanagawa 259-1292, Japan}\\
\addressoftit{Department of Physics, Tokyo Institute for Technology, Meguro, Tokyo 152-8551, Japan }\\
\addressofwarsaw{Institute of Experimental Physics, Warsaw University, 00-681 Warsaw, Poland }\\
\addressofuw{Department of Physics, University of Washington, Seattle, WA 98195-1560, USA}\\
}
\affiliation{ } 


\begin{abstract}
The time variation of the elastic scattering rate of
solar neutrinos with electrons in Super-Kamiokande-I was fit
to the variations expected from active two-neutrino oscillations.
The best fit in the Large Mixing Angle solution has a mixing
angle of $\tasq=0.55$ and a mass squared difference of
$\dmsq=6.3\times10^{-5}$eV$^2$ between the two 
neutrino mass eigenstates. The fitted day/night asymmetry
of $-1.8\pm1.6$(stat)$\plumin{1.3}{1.2}$(syst)\% has improved
statistical precision over previous measurements and agrees well with the
expected asymmetry of -2.1\%.
\end{abstract}

\pacs{14.60.Pq,26.65.+t,96.40.Tv,95.85.Ry}
\maketitle


The combined analysis of all solar neutrino experiments
\cite{solar,skflux} gives firm evidence for neutrino oscillations.
All data are well described using just two neutrino mass eigenstates
and imply a mass squared difference between $\dmsq=3\times10^{-5}$eV$^2$
and $\dmsq=1.9\times10^{-4}$eV$^2$ and a mixing angle between 
$\tasq=0.25$ and $\tasq=0.65$~\cite{global}. This region
of parameter space is referred to as the Large Mixing Angle solution
(LMA). The rate and spectrum of reactor anti-neutrino interactions
in the KamLAND experiment~\cite{kl} are also well reproduced for these
mixing angles and some of these $\dmsq$.
Over the $\dmsq$ range of the LMA, solar $^8$B neutrinos are
$\approx$100\% resonantly converted into the second mass eigenstate
by the large matter density inside the sun~\cite{MSW}.
Therefore, the survival probability into $\nu_e$ is
$\approx\sin^2\theta$. However, due to the presence of the earth's matter
density, the oscillation probability at an experimental site on earth
into $\nu_e$ differs from $\sin^2\theta$ during the night. Since our
experiment is primarily sensitive to $\nu_e$'s, this induces
an apparent dependence of the measured neutrino interaction rate
on the solar zenith angle (often a regeneration of $\nu_e$'s during the night).
We employ a maximum likelihood fit to the expected solar zenith angle
dependence on the neutrino interaction rate. Herein, the statistical uncertainty
is reduced by 25\% compared to our previous measurement of the day/night
asymmetry~\cite{skflux,global} which consists of two flux measurements in
two separate data samples (day and night). It would require almost three
more years of running time to obtain a similar uncertainty reduction
with the previous method.

Super-Kamiokande (SK) is a 50,000 ton water Cherenkov detector described
in detail elsewhere~\cite{skdet}.
SK measures the energy, direction, and time of the
recoil electron from elastic scattering of solar neutrinos with electrons
by detection of the emitted Cherenkov light.
Super-Kamiokande started taking data in April, 1996.
In this report, we analyze the full SK-I low energy data set consisting of
1496 live days
(May $31^{\mbox{st}}$, 1996 through July  $15^{\mbox{th}}$, 2000).

The solar neutrino interactions
are separated from background events by taking advantage of
the strong forward peak of the elastic scattering cross section.
The arrival time of each solar neutrino candidate defines a solar
direction. Using this direction, we calculate the angle
$\theta_{\mbox{\tiny sun}}$ between the reconstructed recoil electron direction
and the solar direction.
The data sample is divided into $N_{\mbox{\tiny bin}}=21$
energy bins:
18 energy bins of 0.5 MeV between 5 and 14 MeV, two energy bins of 1 MeV
between 14 and 16 MeV, and one bin between 16 and 20 MeV.
We use two types of probability density functions:
$p(\cos\theta{\mbox{\tiny sun}},E)$ describes the angular shape expected for solar
$\nu_e$'s of energy $E$ (signal events) and
$u_i(\cos\theta{\mbox{\tiny sun}})$ is the background shape in energy bin $i$.
Each of the 
$n_i$ events in energy bin $i$ is assigned the background factor
$b_{i\kappa}=u_i(\cos\theta_{i\kappa})$ and the signal factor
$s_{i\kappa}=p(\cos\theta_{i\kappa},E_\kappa)$. The likelihood
\[
{\cal L}=e^{-\left(\sum_i B_i+S\right)}\prod_{i=1}^{N_{\mbox{\tiny bin}}}\prod_{\kappa=1}^{n_i}
\left(B_i\cdot b_{i\kappa}+S\frac{\mbox{MC}_i}{\sum_j\mbox{MC}_j}\cdot s_{i\kappa}\right)
\]
is maximized with respect to the signal $S$ and the 21
backgrounds $B_i$. MC$_i$ is the number of events expected
in energy bin $i$ using the flux and spectrum of $^8$B and
{\it hep} neutrinos.

\begin{figure}[tb]
\centerline{\includegraphics[width=8.6cm]{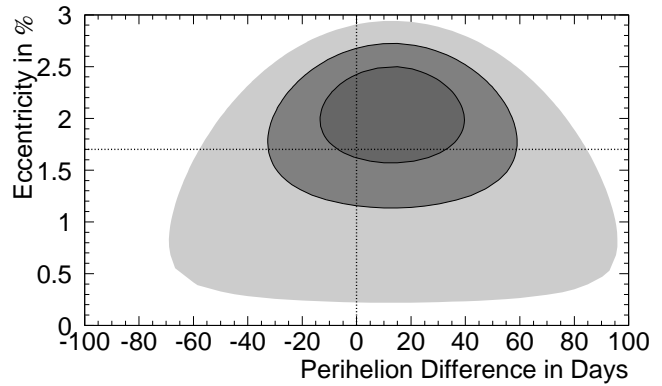}}
\caption{Allowed Regions for the Eccentricity-Induced Seasonal
Solar Neutrino Flux Variation at 68\% (dark gray), 95\% (gray), and
99.73\% (light gray) C.L..}
\label{fig:seasmap}
\end{figure}

A simple determination of the day/night asymmetry 
is obtained by dividing the data sample into day and night and
fit $\cos\theta_{\mbox{\tiny sun}}$ to each sample separately. 
From the obtained day (D) and night (N) rates we calculate
the asymmetry 
$A_{\mbox{\small DN}}=(D-N)/(0.5(D+N))=-2.1\pm2.0$(stat)$\plumin{1.3}{1.2}$(syst)\%
which is consistent with zero.
To take into account time variations in the likelihood fit,
the signal factors are modified
to $s_{i\kappa}=p(\cos\theta_{i\kappa},E_\kappa)\times z_i(\alpha,t_{\kappa})$
where $t_\kappa$ is the event time and $\alpha$ is an
amplitude scaling factor.
As a simple example, we measure the earth's orbital eccentricity.
Since the neutrino flux is proportional to the inverse square of the
distance between sun and earth, the eccentricity induces a seasonal
time variation.
Below $6.5$ MeV the background rates can fluctuate
at time scales of several weeks or longer mainly due to changes in the
radon contamination in water. Therefore, these energy
bins are excluded from the eccentricity analysis by setting $z_i$ to 1.
To measure both the phase
and amplitude of the variation, both the eccentricity and the perihelion
is varied around the known values (1.7\% and $\sim$ January 3rd).
Figure~\ref{fig:seasmap}
shows the allowed ranges of parameters at 68\%, 95\%, and 99.73\% C.L..
We measure the perihelion shift to be $13\pm17$ days (consistent with zero)
and the amplitude of the neutrino flux variation to be
$1.51\pm0.43$ (consistent with one) times
the amplitude expected from 1.7\% eccentricity.

\begin{figure}[tb]
\centerline{\includegraphics[width=8.8cm]{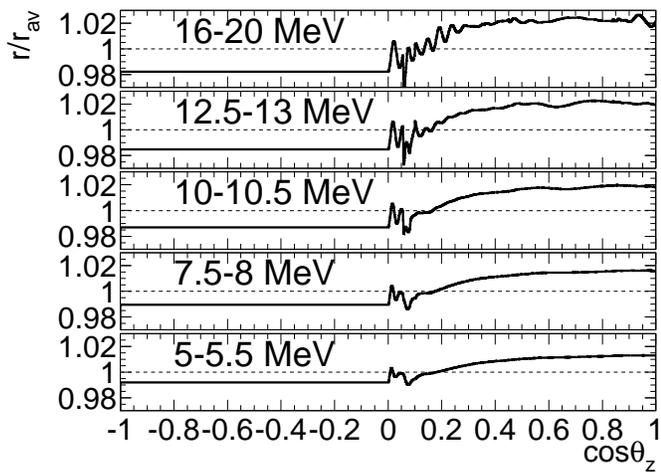}}
\caption{LMA Solar Zenith Angle Variation Shapes.
The predictions are for $\dmsq=6.3\times10^{-5}$eV$^2$
in the energy bins 16 to 20 MeV (top), 12.5 to 13 MeV,
10 to 10.5 MeV, 7.5 to 8 MeV, and 5-5.5 MeV (bottom).}
\label{fig:lmashape}
\end{figure}

To study neutrino
oscillation-induced time variations, we correct for
this seasonal effect (with the nominal perihelion and eccentricity).
After that,
the additional seasonal amplitude variation is $0.48\pm0.43$ times the
eccentricity-induced variation which is consistent with zero.
We search for solar zenith angle variations (employing the solar zenith
angle as the time variable)
and additional seasonal
variation due to the oscillation phase (using the distance between sun
and earth).
In each bin $i$ we calculate the rate $r_i(t)$
(oscillated Monte Carlo).
From this rate and the live-time distribution 
the average ($r^{\mbox{\tiny av}}_i$), day, and night rates
and subsequently the day/night asymmetry $A_i$ are computed.
Using the day (night) live-times $L_D$ ($L_N$) and the
live-time asymmetry $L_{DN}=(L_D-L_N)/(0.5(L_D+L_N))$, the
effective asymmetry parameter $a_i=0.25A_iL_{DN}$ is computed
and $z_i(\alpha,t)$ is defined as
\[
z_i(\alpha,t)=
\frac{1+\alpha\left((1+a_i)r_i(t)/r^{\mbox{\tiny av}}_i-1\right)}{1+\alpha\times a_i},
\]
so that $r^\prime_i(\alpha,t)=z_i(\alpha,t)\times r^{\mbox{\tiny av}}_i$
has the same average total rate $r^{\mbox{\tiny av}}_i$, but the
day/night asymmetry is $A_i\times\alpha$. In particular,
$r^\prime_i(0,t)=r^{\mbox{\tiny av}}_i$ is independent of $t$ and
$r^\prime_i(1,t)=r_i(t)$.
Figure~\ref{fig:lmashape}
shows the expected solar zenith angle variation
shapes $z_i(1,\cos\theta_z)$ in five different energy bins
using an LMA solution and the density model of the earth~\cite{PREM}.

\begin{figure}[tb]
\centerline{\includegraphics[width=8.8cm]{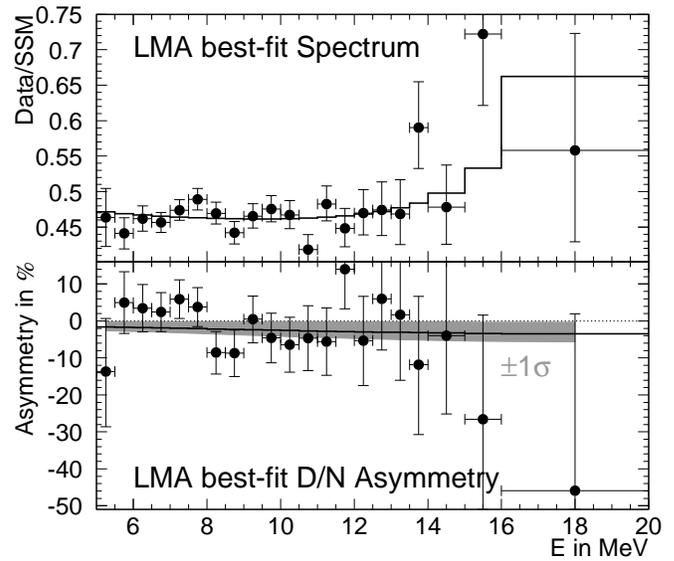}}
\caption{LMA Spectrum (top) and D/N Asymmetry (bottom).
The predictions (solid lines) are for $\tasq=0.55$ and
$\dmsq=6.3\times10^{-5}$eV$^2$ with 
$\phi_{^8B}=0.96\times$Standard Solar Model~\cite{BP2000} and
$\phi_{\mbox{\tiny\em hep}}=3.6\times$Standard Solar Model.
Each energy bin is fit independently to the rate (top)
and the day/night asymmetry (bottom).
The gray bands
are the $\pm1\sigma$ ranges corresponding to the
fitted value over the entire range 5-20 MeV: $A=-1.8\pm1.6$\%.}
\label{fig:lmadnfit}
\end{figure}

\begin{figure}[tb]
\centerline{\includegraphics[width=8.8cm]{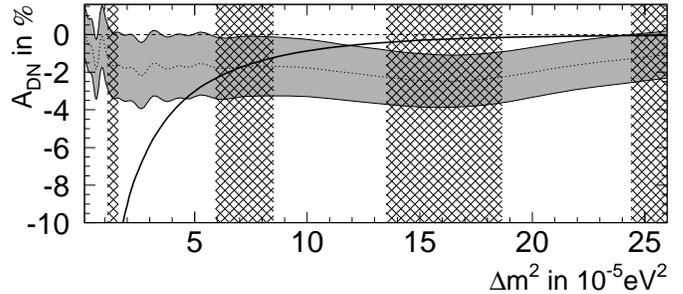}}
\caption{SK Day/Night Asymmetry as a Function of
$\dmsq$. The solid line is expected from two-neutrino oscillations,
the band ($\pm1\sigma$) results from the fit to the SK data.
The mixing angle $\tasq=0.55$ is used. Overlaid are the allowed
ranges in $\dmsq$ (cross-hatched bands) 
from the KamLAND experiment~\cite{kl}. The
second band (LMA-I) is favored.}
\label{fig:dnosc}
\end{figure}

The resulting likelihood function is maximized with respect to
signal $S$, the backgrounds $B_i$, and the asymmetry scaling
parameter $\alpha$. For the best-fit LMA oscillation parameters
(which will be described later)
we find
$\alpha=0.86\pm0.77$ which corresponds to the day/night asymmetry
\[
A_{\mbox{\small DN}}=-1.8\pm1.6\mbox{(stat)}\plumin{1.3}{1.2}\mbox{(syst)}\%
\]
where $-2.1\%$ is expected for these parameters. 
The statistical uncertainty is reduced by 25\%
with this likelihood analysis;
however, the resulting day/night
asymmetry is still consistent with zero.
Figure~\ref{fig:lmadnfit} shows
the fitted rate (top), as well as the day/night asymmetry (bottom)
for each energy bin separately. The oscillation expectations are indicated by
the solid lines. 
The asymmetry fit value and uncertainty depends on the
solar zenith angle variation shapes $z_i(1,t)$ which in turn depend on the
oscillation parameters. Figure~\ref{fig:dnosc} shows the expected
day/night asymmetry and fit results for each $\dmsq$ in the LMA
region with the best-fit mixing angle $\tasq=0.55$.
The expected day/night asymmetry and the $\pm1\sigma$ band
of the fit overlap between $5-12\times10^{-5}$eV$^2$.

To constrain neutrino oscillation using the SK rate time variations,
the likelihood difference 
$\Delta\log {\cal L}=\log {\cal L}(\alpha=1)-\log {\cal L}(\alpha=0)$
between the expected time variation and no time variation is computed.
Below $\dmsq=1.8\times10^{-9}$eV$^2$, the day/night variation is
replaced by an additional seasonal variation due to the oscillation phase.
As for the eccentricity-induced variation, the energy bins below 6.5 MeV
are excluded from the seasonal variation because of the slow time variation
of the background. However, since the effect of that variation on the
day/night asymmetry was carefully evaluated to be negligible, these
energy bins participate in
the day/night variation. To combine the time variation constraints
with those from the recoil electron spectrum, $\Delta\log {\cal L}$
is interpreted
as a time-variation 
$\Delta\chi^2_{\mbox{tv}}=-2\Delta\log {\cal L}$ and added to
the spectrum $\chi^2$.

\begin{figure}[tb]
\centerline{\includegraphics[width=8.8cm]{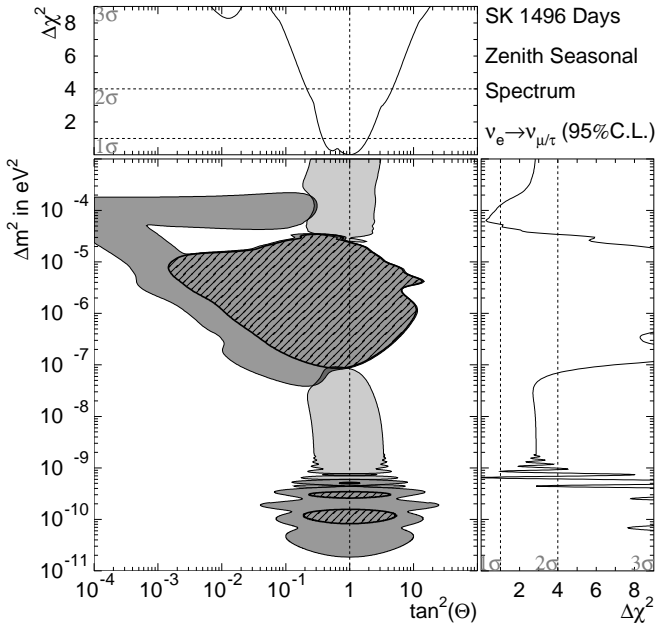}}
\caption{Excluded (SK spectrum and time variation; dark gray)
and Allowed (SK spectrum, rate, and time variation; light gray)
at 95\% C.L.. Overlaid are the areas excluded just by the
day/night and seasonal variation (hatched regions inside thick black lines).
The graphs at the top (and right) show the $\chi^2$ difference
as a function of $\tasq$ ($\dmsq$) alone where
the $\dmsq$ ($\tasq$) is chosen to minimize $\chi^2$.}
\label{fig:skosc}
\end{figure}

Disregarding oscillations, we calculate the interaction
rates in bin $i$ $^8$B$_i$ (hep$_i$) due to $^8$B ({\it hep}) neutrinos.
We also compute the oscillated rates
$^8$B$_i^{\mbox{\tiny osc}}$ and hep$_i^{\mbox{\tiny osc}}$
for each $\tasq$ and $\dmsq$. From these and the measured rates
Data$_i$, we form the ratios
$d_i=$Data$_i$/$(^8$B$_i+$hep$_i)$,
$b_i=^8$B$_i^{\mbox{\tiny osc}}$/$(^8$B$_i+$hep$_i)$, and
$h_i=^8$hep$_i^{\mbox{\tiny osc}}$/$(^8$B$_i+$hep$_i)$.
The expected oscillation suppressions are
$\beta b_i+\eta h_i$ where $\beta$ ($\eta$) is the $^8$B
({\it hep}) neutrino flux scaling parameter. These suppressions are
modified by the correlated uncertainty distortion
functions $f_i^B(\delta_B)$ (uncertainty in the $^8$B neutrino spectrum),
$f_i^S(\delta_S)$ (uncertainty in SK energy scale), and
$f_i^R(\delta_R)$ (uncertainty in SK energy resolution) to
$\rho_i=\frac{\beta b_i+\eta h_i}{f_i}$ where $f_i=f_i^Bf_i^Rf_i^S$.
With the total energy bin-uncorrelated uncertainty
$\sigma_i^2=\sigma_{i,\mbox{\tiny stat}}^2+\sigma_{i,\mbox{\tiny sys,u}}^2$
the total $\chi^2$ is then
\[
\chi^2=\sum_{i=1}^{N_{\mbox{\tiny bin}}}\left(\frac{d_i-\rho_i}{\sigma_i}\right)^2
+\frac{\delta_B^2}{\sigma_B^2}+\frac{\delta_S^2}{\sigma_S^2}
+\frac{\delta_R^2}{\sigma_R^2}+\Delta\chi^2_{\mbox{tv}}
+\left(\frac{\beta-1}{\sigma_f}\right)^2
\]
where the last term constraining the $^8$B flux
to the standard solar model~(SSM)~\cite{BP2000} is optional.
Including this last term, the best oscillation fit
is in the quasi-vacuum region at
$\dmsq=6.49\times10^{-10}$eV$^2$ and maximal mixing,
where a summer/winter asymmetry
of -0.6\% is expected and $-0.3\pm0.7\%$(stat) is found. The
$\chi^2$ is 17.1 for 20 degrees of freedom (65\% C.L.).
The LMA solution fits almost equally well: the smallest $\chi^2$
at $\dmsq=6.3\times10^{-5}$eV$^2$ and
$\tasq=0.55$ is 17.3 (63\% C.L.). Figure~\ref{fig:skosc} shows
the allowed areas at 95\% C.L. using all SK information:
rate, spectrum and time-variation. It also shows the
$\Delta\chi^2$ as a function of $\tasq$ ($\dmsq$) alone:
SK data excludes small mixing at more than $3\sigma$.
SK data also disfavors $\dmsq>10^{-3}$eV$^2$ and
$2\times10^{-9}$eV$^2<\dmsq<3\times10^{-5}$eV$^2$.

\begin{figure}[tb]
\centerline{\includegraphics[width=8.8cm]{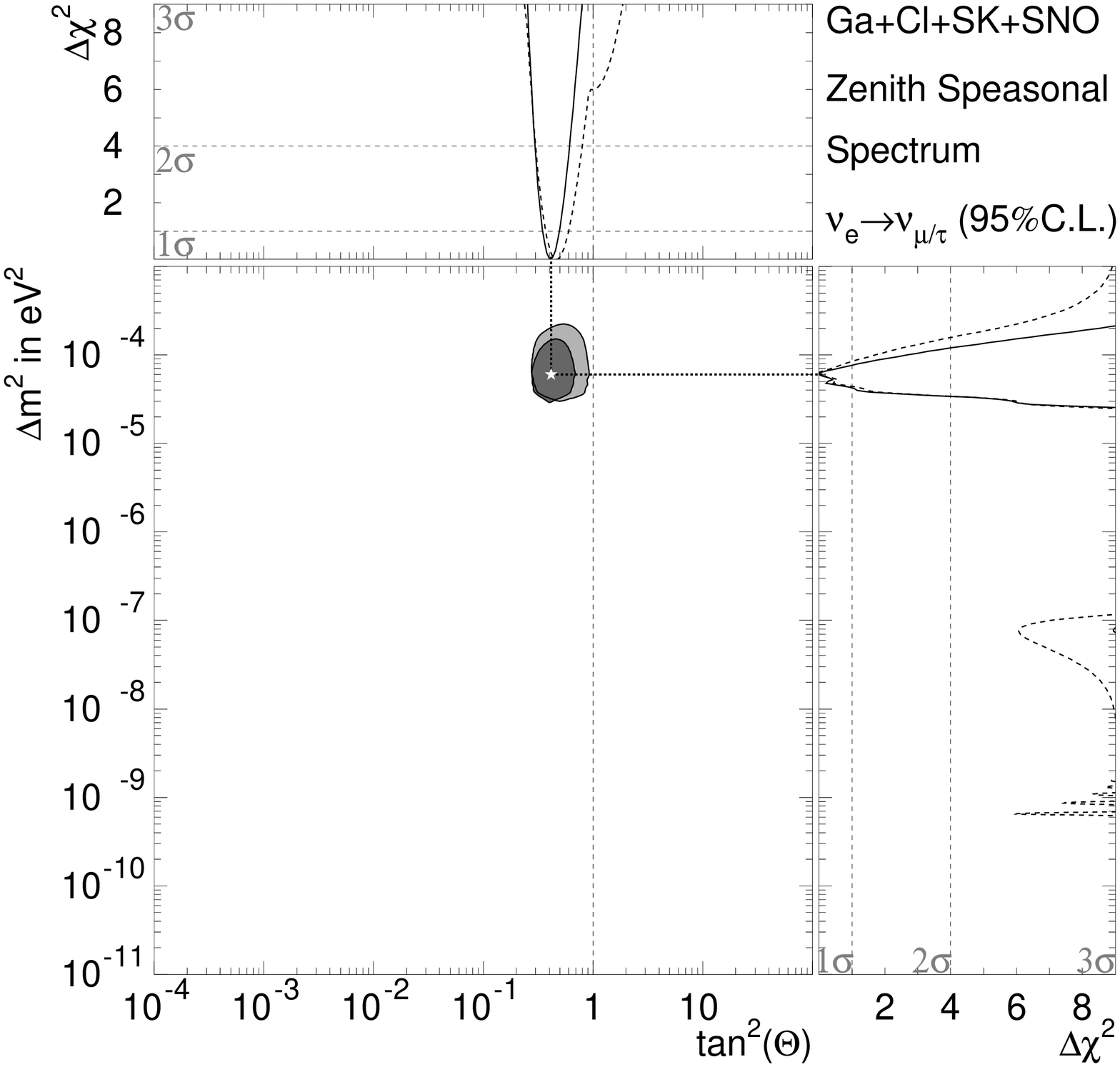}}
\caption{Allowed area at 95\% C.L from the combination of
SK and SNO (gray) and all solar data (dark gray).
The graphs at the top (and right) show the $\chi^2$ difference
as a function of $\tasq$ ($\dmsq$) only: the dashed line is
the SK/SNO fit, the solid line includes all solar data.
The best fit to all solar data is
$\tasq=0.42$ and $\dmsq=6.0\times10^{-5}$eV$^2$.}
\label{fig:globosc}
\end{figure}

Stronger constraints on $\dmsq$ result from the combination of
SK with other solar neutrino data~\cite{solar}. The combined
fit to SK data and the SNO measurements on the charged-current
and neutral-current reactions of solar $^8$B neutrinos with
deuterons need not constrain any neutrino flux with a solar model.
Figure~\ref{fig:globosc} shows the allowed region at 95\%C.L.:
only LMA solutions survive.
When the charged-current rates measured by Homestake, GALLEX/GNO,
and SAGE are included as well, the LMA solutions are favored
by $3\sigma$; however, the fit relies on the SSM predictions of the
pp, pep, CNO, and $^7$Be neutrino fluxes.

\begin{figure}[tb]
\centerline{\includegraphics[width=8.5cm]{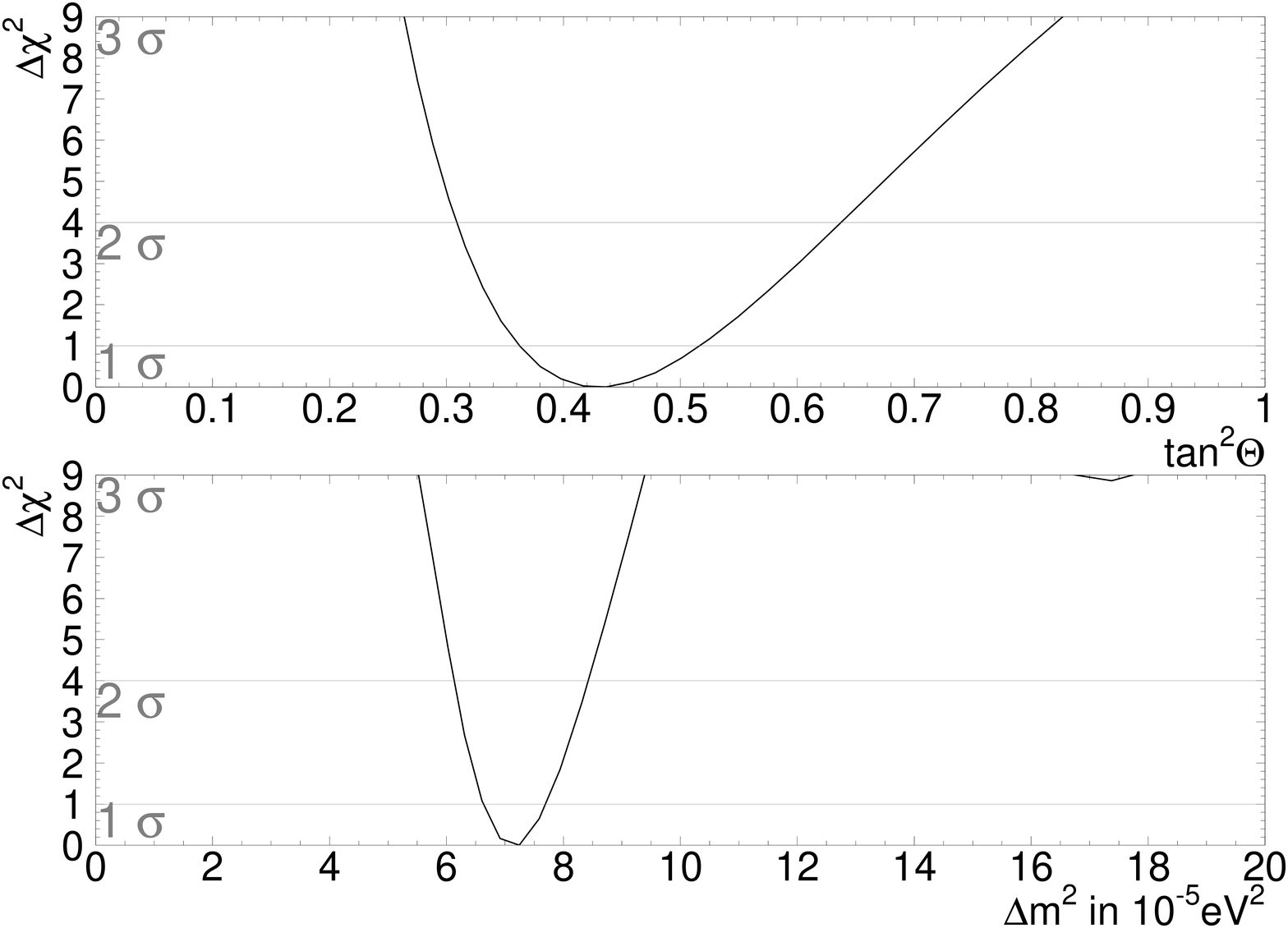}}
\caption{$\Delta\chi^2$ Curves as a Function of Mixing (Top)
and $\dmsq$ (Bottom) Using all Solar and KamLAND Data.
Only LMA-I remains allowed at $3\sigma$.}
\label{fig:globkl}
\end{figure}

The first oscillation analysis of the KamLAND reactor neutrino spectrum
and rate leaves several allowed areas, usually called LMA-0,
LMA-I, LMA-II, and LMA-III. When we combine the analysis of SK
with all other solar experiments and a likelihood analysis of
the KamLAND data~\cite{ianni},
the LMA-I is strongly favored over the other solutions:
in Figure~\ref{fig:globkl}, the $\Delta\chi^2$ of the fit is plotted
against $\dmsq$ and $\tasq$.

In summary, SK has measured very precisely the $^8$B neutrino
flux time-variations expected from two-neutrino oscillations.
For the best LMA parameters, the day/night asymmetry is determined
as $=-1.8\pm1.6$(stat)$\plumin{1.3}{1.2}$(syst)\% where
$-2.1\%$ is expected. SK data disfavor large 
$\Delta m^2$ LMA solutions, since their expected day/night asymmetries
are closer to zero.
In combination with other solar data and the KamLAND reactor neutrino
results,
the oscillation parameters are determined as
$\dmsq=7.1\plumin{0.6}{0.5}\times10^{-5}$eV$^2$
and $\tasq=0.44\pm0.08$.

The authors acknowledge the cooperation of the Kamioka Mining and
Smelting Company.  The Super-Kamiokande detector has been built and
operated from funding by the Japanese Ministry of Education, Culture,
Sports, Science and Technology; the U.S. Department of Energy; and the
U.S. National Science Foundation.  This work was partially supported
by the Korean Research Foundation (BK21) and the Korea Ministry of
Science and Technology.


\begin{thebibliography}{99}

%
%
\addtocounter{notes}{1}
\bibitem[\fnsymbol{notes}]{aaa}
Present address: Harvard University, Cambridge, MA 02138, USA


\addtocounter{notes}{1}
\bibitem[\fnsymbol{notes}]{ccc}
Present address: The Institute of Physical and Chemical Reasearch (RIKEN), Wako, Saitama 351-0198, Japan

\addtocounter{notes}{1}
\bibitem[\fnsymbol{notes}]{ddd}
Present address: Department of Physics, University of Utah, Salt Lake City, UT 84112, USA


\bibitem{solar} B.T.Cleveland et al.,         \Journal{\APJ}{496}{505}{1998};
                V.Gavrin,                     \Journal{\NPBs}{118}{39}{2003};
                T.Kirsten,                    \Journal{\NPBs}{118}{33}{2003};
                S.Fukuda et al.,              \Journal{\PRL}{86}{5656}{2001};
                Q.R.Ahmad et al.,             \Journal{\PRL}{89}{011301}{2002};
		Q.R.Ahmad et al.,             \Journal{\PRL}{89}{011302}{2002}.
\bibitem{skflux}S.Fukuda et al.,              \Journal{\PRL}{86}{5651}{2001}.
\bibitem{global}S.Fukuda et al.,              \Journal{\PLB}{539}{179}{2002}.
\bibitem{kl}    K.Eguchi et al.,              \Journal{\PRL}{90}{021802}{2003}.
\bibitem{MSW}   S.P.Mikheyev and A.Y.Smirnov, \Journal{\SNP}{42}{913}{1985};
                L.Wolfenstein,                \Journal{\PRD}{17}{2369}{1978}.
\bibitem{skdet} S.Fukuda et al.,              \Journal{\NIMA}{501}{418}{2003}.
\bibitem{PREM} A.M.Dziewonski and D.L.Anderson, Phys. Earth Planet. Inter. {\bf 25}, 297 (1981).
\bibitem{BP2000}J.N.Bahcall et al.,           \Journal{\APJ}{555}{990}{2001}.
\bibitem{ianni} A.Ianni, {\it hep-ph/0302230v2; to be published in Journal of Physics G}
\end{thebibliography}
\end{document}